\documentclass[12pt]{iopart}
\usepackage{graphicx}
\usepackage{epstopdf}
\usepackage{bm}
\begin{document}
\title{Pressure Engineering of the Dirac Fermions in Quasi-One-Dimensional Tl$_2$Mo$_6$Se$_6$}

\author{Ziwan Song$^1$, Bin Li$^{2,3}$, Chunqiang Xu$^{4,5}$, Sixuan Wu$^6$, Bin Qian$^5$, Tong Chen$^7$, Pabitra K. Biswas$^8$, Xiaofeng Xu$^{4,5}$, Jian Sun$^7$}
\address{$^1$ College of Electronic and Optical Engineering, Nanjing University of Posts and Telecommunications, Nanjing 210023, China}
\address{$^2$ New Energy Technology Engineering Laboratory of Jiangsu Province and School of Science, Nanjing University of Posts and Telecommunications, Nanjing 210023, China}
\address{$^3$ National Laboratory of Solid State Microstructures, Nanjing University, Nanjing 210093, China}
\address{$^4$ Department of Applied Physics, Zhejiang University of Technology, Hangzhou 310023, China}
\address{$^5$ Department of Physics, Changshu Institute of Technology, Changshu 215500, China}
\address{$^6$ School of Science, Nanjing University of Posts and Telecommunications, Nanjing 210023, China}
\address{$^7$ National Laboratory of Solid State Microstructures and School of Physics, Nanjing University, Nanjing 210093, China}
\address{$^8$ ISIS Pulsed Neutron and Muon Source, STFC Rutherford Appleton Laboratory, Harwell Campus, Didcot, Oxfordshire OX11 0QX, United Kingdom}
\ead{libin@njupt.edu.cn (B Li) and {xiaofeng.xu@cslg.edu.cn} (X Xu)}


\begin{abstract}

Topological band dispersions other than the standard Dirac or Weyl fermions have garnered the increasing interest in materials science. Among them, the cubic Dirac fermions were recently proposed in the family of quasi-one-dimensional conductors A$_2$Mo$_6$X$_6$ (A= Na, K, In, Tl; X= S, Se, Te), where the band crossing is characterized by a linear dispersion in one $k$-space direction but the cubic dispersion in the plane perpendicular to it. It is not yet clear, however, how the external perturbations can alter these nontrivial carriers and ultimately induce a new distinct quantum phase. Here we study the evolution of Dirac fermions, in particular the cubic Dirac crossing, under external pressure in the representative quasi-one-dimensional Tl$_2$Mo$_6$Se$_6$ via the first-principles calculations. Specifically, it is found that the topological properties, including the bulk Dirac crossings and the topological surface states, change progressively under pressure up to 50 GPa where it undergoes a structural transition from the hexagonal phase to body-centered tetragonal phase. Above 50 GPa, the system is more likely to be topologically trivial. Further, we also investigate its phonon spectra, which reveals a gradual depletion of the negative phonon modes with pressure, consistent with the more three-dimensional Fermi surface in the high-pressure phase. Our work may provide a useful guideline for further experimental search and the band engineering of the topologically nontrivial fermions in this intriguing state of matter.

\end{abstract}
\noindent{\bf Keywords:}
\noindent{\it Cubic Dirac Fermions, Topological Superconductivity, Quasi-one-dimensional Superconductors\/}\\
\submitto{JPCM}
\maketitle

\section{Introduction}

The search for new topological states has become the cynosure of condensed matter physics since the discovery of topological insulators \cite{Hasan_RMP,Zhang_RMP}. With the advent of Dirac and Weyl semimetals (DSMs and WSMs), the research interest in topological phenomena has largely shifted towards various breeds of topological metals or semimetals, including the topological nodal-line semimetals and those with fermionic excitations beyond 4-fold (Dirac) or 2-fold (Weyl) degeneracies\cite{Zhou19}, e.g., the threefold/sixfold/eightfold degenerate point nodes\cite{Bernevig16,Zhang17}. More recently, a new type of topological fermion has been theoretically proposed, namely the cubically dispersed Dirac semimetal (CDSM)\cite{Gibson2015,Liu2017}. The CDSMs possess linear band crossing along one principle axis but the cubic dispersions in the plane perpendicular to it, i.e., $E(k)\propto k^n$ ($n$=3), $k$ is the wave vector measured with respect to the crossing node\cite{Liu2017}. According to the different \textit{in-plane} dispersion power $n$ at the crossing point, Dirac/Weyl points can be classified as linear ($n$=1), quadratic ($n$=2) and cubic ($n$=3). The material incarnations of this CDSM are extremely rare due to the severe crystal symmetry constraints, with only the quasi-one-dimensional (q-1D) molybdenum chalcogenides A$_2$Mo$_6$X$_6$ (A= Na, K, Rb, In, Tl; X= S, Se, Te) being the leading candidates to realize these cubic Dirac fermions\cite{Liu2017}.

Q-1D A$_2$Mo$_6$X$_6$ systems were first discovered in the early 1980s and attracted immense interest primarily due to their one-dimensionality\cite{Potel1980,Armici1980,Huang1983,Tarascon1984,Hor85,Hor85_SSC,Brusetti1988,Tessema91,Brusetti94,Petrovic2010}. Contrary to the higher-dimensional counterparts, interaction plays a more pronounced role in q-1D systems since electrons confined in one dimensionality can barely avoid the effects of interactions. Due to the strong interactions, the thermal and quantum fluctuations are significant, and prevent the breaking of continuous symmetries. Consequently, q-1D A$_2$Mo$_6$X$_6$ provides a distinctive playground for studying the emergent phenomena in low-dimensional physics, including non-Fermi liquid behaviors, charge density wave, and topological superconductivity. Understanding the physics of q-1D electrons has been the focus of extensive theoretical and experimental efforts (for a review, see Ref. \cite{Giamarchi_2004}). Structurally, these A$_2$Mo$_6$X$_6$ materials are composed of one-dimensional (Mo$_6$Se$_6$)$_\infty$ chains oriented along the $c$-axis, weakly coupled by the $A$ ions. As a result, the interchain resistivity was found to be 3 orders of magnitude larger than that of intrachain resistivity, making it among the most anisotropic materials known to exist. Interestingly, some members in this family, like Tl$_2$Mo$_6$Se$_6$, Na$_2$Mo$_6$Se$_6$ and In$_2$Mo$_6$Se$_6$, are also superconducting at ambient pressure. Remarkably, a novel disorder-enhanced superconductivity was reported in Na$_2$Mo$_6$Se$_6$, suggesting the disorder-induced Coulomb pair-breaking effect being averted due to the screened long-range Coulomb repulsion by disorder\cite{Petrovic2016}. More recently, theoretical works reveal the coexistence of quadratic and cubic Dirac fermions, in addition to the conventional linear Dirac point, in this series of q-1D compounds\cite{Liu2017}.

As a notable example in this family, Tl$_2$Mo$_6$Se$_6$ becomes superconducting with $T_c$ varying between 3 and 6.5 K, depending on the stoichiometry. More strikingly, topological superconductivity was recently proposed in Tl$_2$Mo$_6$Se$_6$ by theory, sparking renewed interest in these materials\cite{Gannon18,Chia18,Huang18}. The questions may arise, however, as to how robust these Dirac fermions in Tl$_2$Mo$_6$Se$_6$ are to the variations in the lattice parameters or in its one-dimensionality, and how fragile these topological properties are to other instabilities, such as charge-density waves that are often observed in q-1D materials. All these questions invoke more studies on this newly proposed topological material, both theoretically and experimentally. On the other hand, it is well established that pressure can effectively modify the atomic and electronic structure of a material, giving rise to the novel phases with unusual physical properties, such as a new superconducting phase. In this sense, it is highly desirable to study the topological evolution of the band structure of Tl$_2$Mo$_6$Se$_6$ under pressure and search for new possible phases in it.

In this work, by means of the first-principles calculations, we study the pressure evolution of the band structure of Tl$_2$Mo$_6$Se$_6$ up to 150 GPa, with special focus on its topological properties. We identify a possible topological phase transition within the $P6_3/m$ structure below 50 GPa. Above 50 GPa, a structural phase transition takes place which drives the system into a topologically trivial phase. The bulk-boundary correspondence has also been studied which explicitly reveals a dramatic modification of the surface states under pressure. The calculated phonon spectra shed light on the possible Peierls distortion in this system. The prediction made in this work shall open up avenues for further study of novel physical properties associated with this newly proposed q-1D topological CDSM. This work may also stimulate the study of cubic Dirac dispersion in other areas of physics, including topological phases of excitations such as magnons and polaritons\cite{Downing_2019,Pirmoradian_2018}.

\section{Methods}
We employed the WIEN2K code~\cite{Wien2k}  with generalized gradient approximation (GGA)~\cite{GGA} to calculate the electronic band structures. The muffin tin radii were chosen to be 2.5 a.u.\ for Tl, and 2.33 a.u.\ for Mo and Se.  A tight-binding model based on Wannier functions~\cite{wannier1} was constructed to obtain the topological properties, using Tl $s$ and $p$, Mo $d$, and Se $p$ orbitals with spin-orbit coupling (SOC) included. The surface states spectrum were calculated with the surface Greens functions as implemented in the WannierTools code~\cite{wannier2}.  We used the Quantum-ESPRESSO program~\cite{QE} to perform the calculations of phonon spectra and electron-phonon couplings. The cutoffs were chosen as 50 Ry for the wave functions and 400 Ry for the charge density. Furthermore, we employed the evolutionary crystal structure prediction method USPEX~\cite{ux1,ux2,ux3} to determine the high-pressure structures.

\section{Results}

In this section, we shall describe how the application of the external pressure affect the bulk electronic structure, phonon stability and topological surface states of Tl$_2$Mo$_6$Se$_6$ in turn, followed by a structure prediction at higher pressures. At ambient pressure, Tl$_2$Mo$_6$Se$_6$ crystallizes in the hexagonal $P6_3/m$ space group as reported previously~\cite{Potel1980}. The crystallographic structure of Tl$_2$Mo$_6$Se$_6$ and the corresponding Brillouin zone with high-symmetry $k$-points are illustrated in Fig.\ 1. The crystalline structure consists of a one-dimensional condensation of an infinite number of Mo$_6$Te$_6$ units running along the $c$-axis, separated by the monovalent Tl$^{+}$ cations. The face-sharing Mo$_6$ octahedra can be viewed as Mo$_3$ triangles related to each other by a screw axis; as such, the crystal structure is nonsymmorphic. Each Mo$_3$ triangle has one extra electron donated by the Tl$^{+}$ ion.

\begin{figure*}
\begin{center}
\includegraphics[width=12cm]{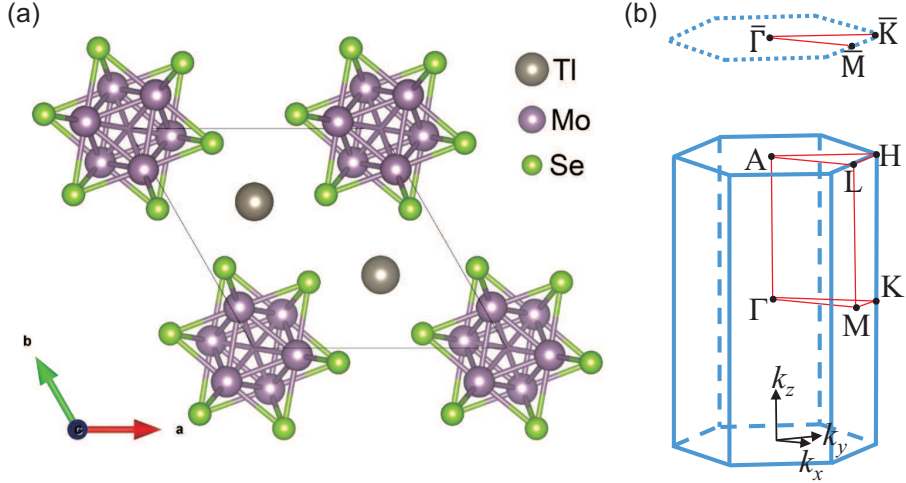}
\caption{(a) The top view of crystallographic structures of Tl$_2$Mo$_6$Se$_6$ in the $P6_3/m$ phase. (b) The Brillouin zone with high-symmetry $k$-points: $\Gamma$(0 0 0), M(0 0.5 0), K(1/3 1/3 0), A(0 0 0.5), H(1/3 1/3 0.5), L(0 0.5 0.5). The dotted lines represent the projected surface Brillouin zone. }
\label{struct}
\end{center}
\end{figure*}

\begin{figure*}
\begin{center}
\includegraphics[width=12cm]{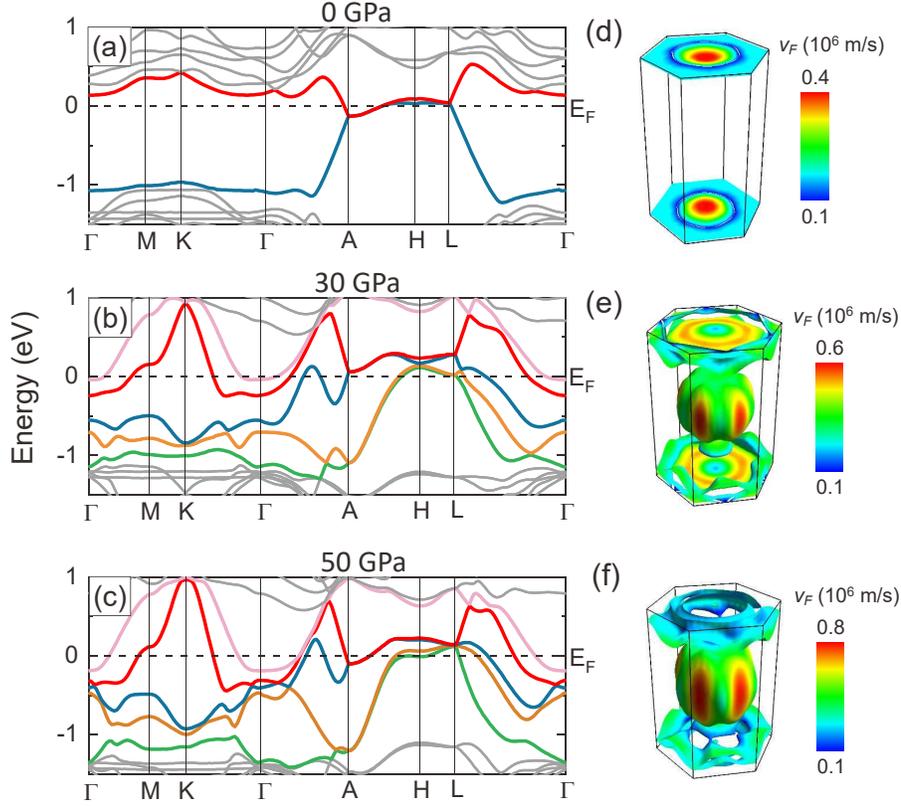}
\caption{Electronic structures of $P6_3/m$ phase under a pressure of (a) 0 GPa, (b) 30 GPa, and (c) 50 GPa. The bands crossing the Fermi level are marked by different colors. The corresponding Fermi surfaces are shown in (d-f) shaded by the Fermi velocity $v_F$.} \label{bands}
\end{center}
\end{figure*}

We now focus on the pressure evolution of the band structure, with close attention to its Dirac crossing points. The calculated bulk band structure and Fermi surface with SOC under three representative pressures (0, 30, 50 GPa) are shown in Fig. \ref{bands}. Our results for the ambient pressure (Fig. \ref{bands}(a)) are in good agreement with previous reports~\cite{Gibson2015,Liu2017}. There are two bands crossing the Fermi level. The bands along the $c$ axis ($\Gamma$-A and $\Gamma$-L) are highly dispersive. In contrast, the dispersions along the $\Gamma$-M-K-$\Gamma$ ($k_z$ = 0 plane) and A-H-L ($k_z = \pi$ plane) are relatively flat, reflecting the q-1D feature of the structure. The SOC splitting is clearly visible between A and L points. Within the energy window of $\sim$1 eV around the Fermi level, there are only two linear bands from conduction and valence bands crossing at A and L points from $\Gamma$. As extensively discussed in Ref. ~\cite{Liu2017}, the band crossing at A has a linear dispersion along $\Gamma$-A and becomes cubically dispersed in the perpendicular directions, e.g., along A-H, thus confirming the cubic Dirac fermions at A. By contrast, the Dirac point at L has linear dispersions along all directions in momentum space albeit with highly anisotropic velocities, consistent with an anisotropic Dirac cone. It should be remarked that there are possibly more Dirac-like crossings in the path of $\Gamma$-A at 300-500 meV above the Fermi level. However, these Dirac-like points are beyond the scope of our current interest in this study. Furthermore, the Fermi surface analysis has been performed to determine the band type using SKEAF (Supercell K-space Extremal Area Finder)~\cite{SKEAF}. It is shown that Mo-derived 4$d$ orbital dominates at the Fermi level and the blue dispersion in Fig. \ref{bands}a is hole-like while the red one is electron-like. The bulk Fermi surface with color-coded Fermi velocity are visualized in the right panel (Fig. \ref{bands}d), which is composed of two parts: the outer hole-like FSs and the inner electron-like FSs near the top and bottom boundaries of the first Brillouin zone. The flatness of the FSs also indicates the q-1D feature at ambient pressure. As the pressure increases, however, the band dispersion is dramatically reconstructed (see Fig. \ref{bands}b and c). Distinctly different from the ambient case, there are totally five bands crossing the Fermi level for 30 and 50 GPa. At 30 GPa, the original Dirac bands shifts upward at L point and a new Dirac point arises at the Fermi level from the original valence bands, resulting in two Dirac crossings at the same $k$-point separated by $\sim$200 meV in energy. At 50 GPa, this new Dirac point shifts further up in energy and touches with the original Dirac point, making a fourfold degenerate point at L. On the other hand, the cubic Dirac point at A is largely unchanged up to 50 GPa except that it moves up or down in energy. As for the Fermi surface, the original flat q-1D FSs become more three-dimensional under pressure, especially in the Brillouin zone center where a barrel-like electron pocket emerges at 30 GPa and 50 GPa. This is conceivable because the dissociated Tl atoms at zero pressure become bonded with their neighboring Se ions and form the Tl-Se$_8$ cuboid with pressure. Meanwhile, the face-sharing Mo$_6$ octahedra collapse at high pressure. As a consequence, the dramatic change in the bonding induced by pressure causes the reconstruction of charge distribution, band dispersion and Fermi surface morphology.

\begin{figure*}
\begin{center}
\includegraphics[width=12cm]{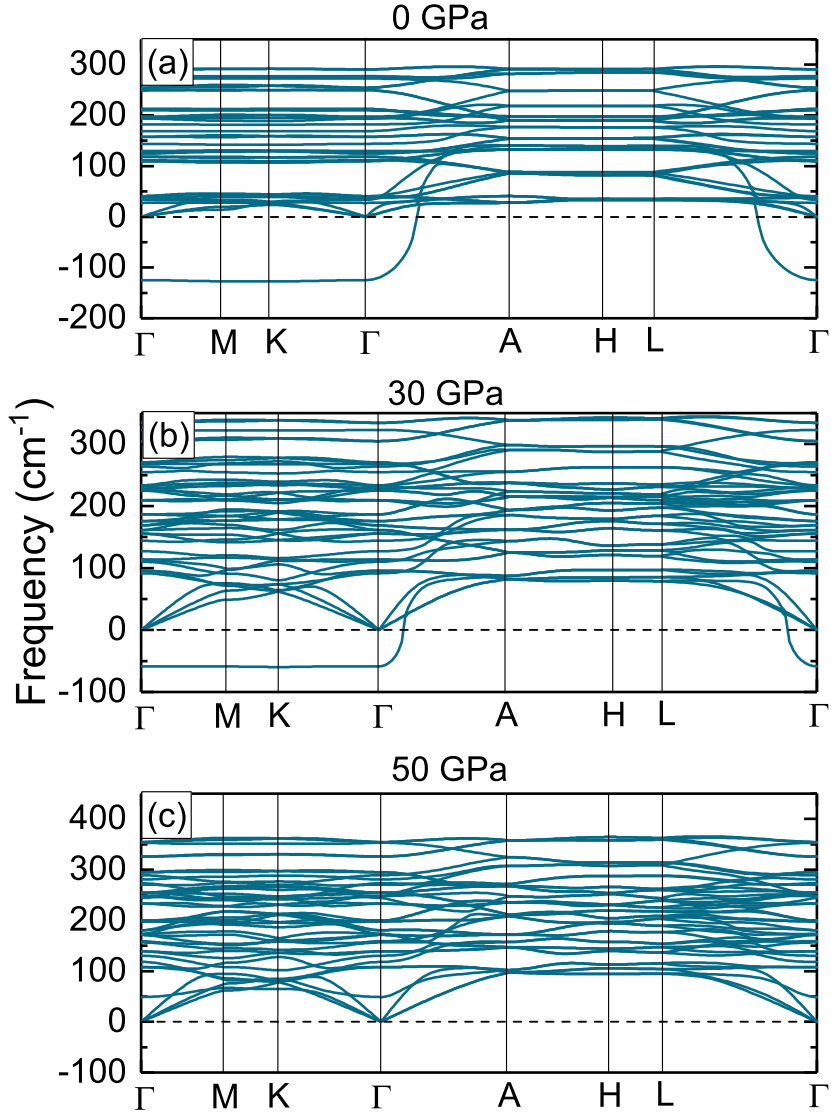}
\caption{Calculated phonon dispersions at (a) ambient pressure, (b) 30 GPa, and (c) 50 GPa.} \label{phonon}
\end{center}
\end{figure*}

Figure 3 depicts the phonon spectra of Tl$_2$Mo$_6$Se$_6$ under 0 GPa, 30 GPa and 50 GPa (all with $P6_3/m$ space group). At 0 GPa, soft phonon modes are found in the $k_z = 0$ plane ($\Gamma$-M-K-$\Gamma$) as well as in the directions of $\Gamma$-A and $\Gamma$-L, indicating that such a high-symmetry semimetallic structure is dynamically unstable against density wave formation. In one-dimensional systems, Peierls demonstrated that at low temperature an instability can be induced by the coupling between carriers and a periodic lattice. Such an instability induces a charge ordering phenomenon and a metal-insulator phase transition~\cite{Peierls}. Such an electron-phonon-driven transition is expected to be second order. If one applies a distorted structure with a reduced symmetry of space group $P\overline{3}$, the negative phonon modes would naturally be depleted with an electronic band gap opened (semimetal-insulator transition)~\cite{Liu2017}. Alternatively, as we apply a pressure to the undistorted structure, the negative phonon modes would be progressively eliminated. As shown in Fig. \ref{phonon}c at 50 GPa, no soft phonon mode can be observed. This is consistent with the experimental finding that pressure can gradually tune this family of materials away from a metal-insulator transition, making them more metallic and in some extreme cases, becoming superconducting\cite{Huang1983,Hor85_SSC,Hor85}.

\begin{figure*}
\begin{center}
\includegraphics[width=12cm]{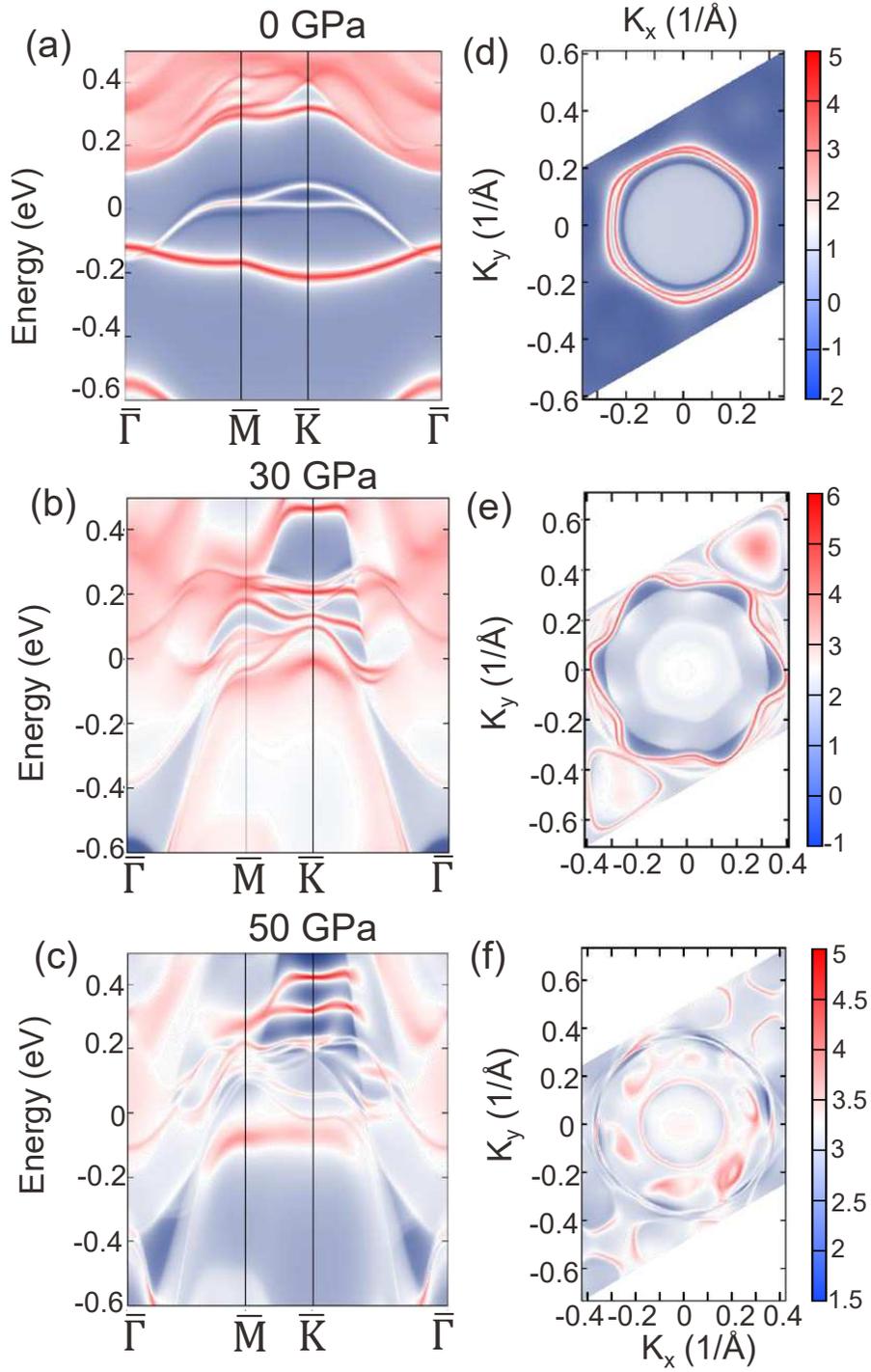}
\caption{\label{surfaces} Calculated surface band structures at (a) 0 GPa, (b) 30 GPa, (c) 50 GPa; (d-f) The corresponding surface state spectra on the (0 0 1) plane. The red dots in (d) and (e) are the projection of the bulk Dirac points L. Bright red lines denote the surface states.}
\end{center}
\end{figure*}

The bulk-boundary correspondence in topology prompts us to study its surface states under pressure. We calculate the surface band structure on the (0 0 1) plane using a tight-binding model based on maximally localized Wannier functions, as shown in Fig. \ref{surfaces}. As the Dirac point in a DSM could be understood as two degenerate Weyl points with opposite chirality, one might expect that there are two copies of Fermi arcs on the DSM surface, forming a ring with two singularities at the surface projection of the Dirac points in the bulk\cite{Potter_NatComm,Hasan_Science_Na3Bi,Hasan_Science,Huang_NatComm,Ding_NatPhys}. Recently, however, Kargarian \textit{et al.} argued that the Fermi arcs on the DSM surface are not topologically protected~\cite{Kargarian} and can be continuously deformed into a closed Fermi coutour without any symmetry breaking. From Fig. \ref{surfaces}(d), one can clearly see the closed Fermi contour surrounding the $\Gamma$ point, instead of two Fermi arcs connecting the projection of the bulk Dirac nodes. Under 30 GPa, the Fermi contour on DSM surface is continuously deformed into the one shown in Fig. \ref{surfaces}(e) without any symmetry breaking. At 50 GPa, only few surface states survived, which denotes weak topological character. The calculated $Z_2$ index suggests the possible topological phase transitions under pressure within this $P6_3/m$ phase: $Z_2$ index is (1, 010) at zero pressure; At 30 GPa, the calculated $Z_2$ invariant numbers are 1 for $k_x$ = 0, $k_y$ = 0, and $k_z$ = $\pi$ plane, and zeros for other planes, thus the topological index is (1, 001), which indicates a strong topological material; At 50 GPa, only few surface states survive with topological index (0, 001), which denotes weak topological character. As a result, in the $P6_3/m$ phase, Tl$_2$Mo$_6$Se$_6$ may undergo a second order topological phase transition from a DSM to a strong topological metal and then to a weak topological metal.

\begin{figure*}
\begin{center}
\includegraphics[width=12cm]{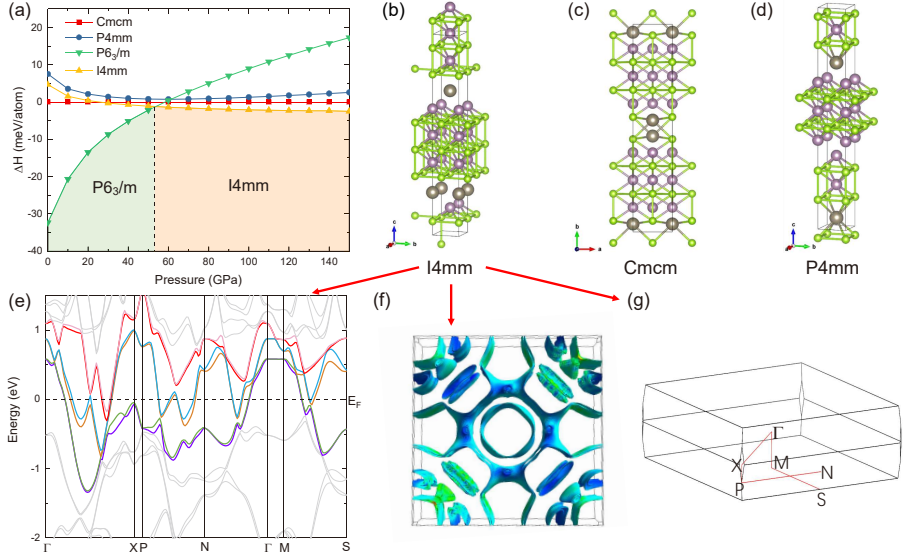}
\caption{\label{enthalpy} (a) Calculated enthalpy curves for $P4mm$ (circles), $P6_3/m$ (down-triangles) and $I4mm$ (up-triangles) phases with respect to the $Cmcm$ structure (squares), as a function of pressure from 0 to 150 GPa. Crystal structure of (b) $I4mm$, (c) $Cmcm$, and (d) $P4mm$. (e-g) The band structure, top view of Fermi surfaces and the Brillouin zone of the $I4mm$ structure at 80 GPa, respectively. The bands crossing the Fermi level in (e) are marked by different colors.}
\end{center}
\end{figure*}

We further use crystal structural prediction techniques to find energetically stable structures of Tl$_2$Mo$_6$Se$_6$ under extremely high pressures. The calculated enthalpy-pressure ($\Delta H$-$P$) curves are plotted in Figs. \ref{enthalpy}(a) for the structures of interest. Several crystallographic structures shown in Fig. \ref{enthalpy} are found to be the stable/metastable phases with lower energies at high pressures. In our calculations, we find three candidate structures under high pressures: orthorhombic $Cmcm$ phase, tetragonal $P4mm$ phase and body-centered tetragonal $I4mm$ phase. Among these candidates, body-centered tetragonal phase has the lowest enthalpy  when the pressure is larger than 50 GPa. Thus, with increasing pressure, Tl$_2$Mo$_6$Se$_6$ may undergo a structural phase transition, from the hexagonal $P6_3/m$ to $I4mm$ structure at about 50 GPa. The unit cell of $I4mm$ structure consists of one formula unit, containing edge-sharing Mo-Se$_{8}$ cubes. Mo ions are located at the center of the cube and Tl ions stay in the space between the cubes. The calculated band structure, Fermi surfaces and Brillouin zone with high-symmetry $k$ points are shown in Fig. \ref{enthalpy}(e-g) for the $I4mm$ structure at 80 GPa. As noted, there are six bands crossing the Fermi level, constructing the complicated three dimensional Fermi surfaces. The further topological band analysis shows that the $I4mm$ structure is a topological trivial phase.

\begin{figure*}
\begin{center}
\includegraphics[width=12cm]{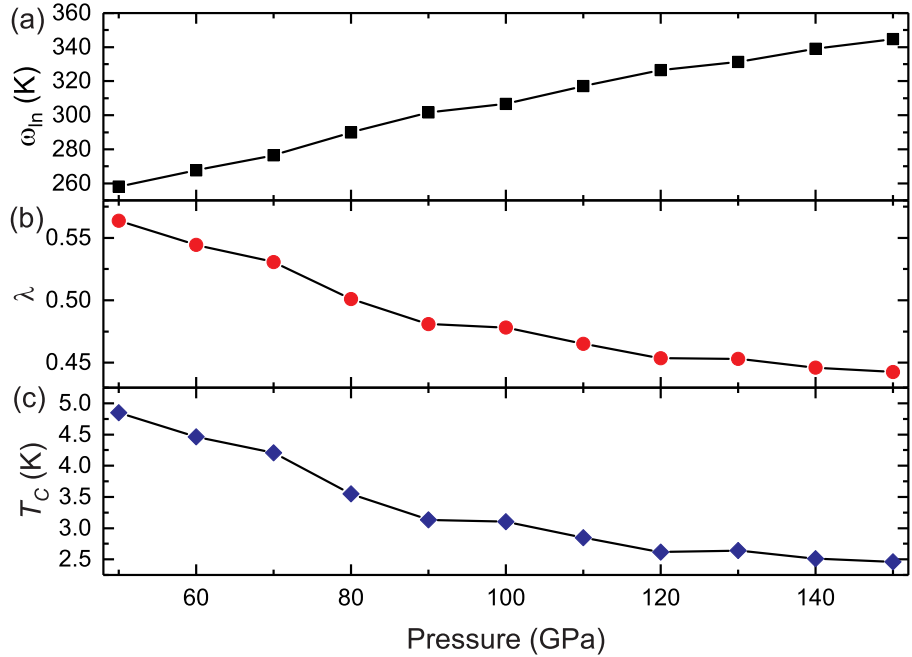}
\caption{\label{Tc} (a) The logarithmically averaged phonon frequency $\omega_{ln}$, (b) electron-phonon coupling constant $\lambda$, and (c) superconducting $T_{c}$ as a function of pressure in the $I4mm$ structure. }
\end{center}
\end{figure*}

In order to estimate the superconducting $T_c$ of Tl$_2$Mo$_6$Se$_6$ under pressure, we performed the linear response calculations of its electron-phonon properties, and estimated the critical temperature through the Mc-Millan Allen-Dynes formula~\cite{Mcmillan1968,Allen1975}

{
\begin{equation}
k_{B}T_{c}=\frac{\hbar\omega_{ln}}{1.2}\textrm{exp}\left[-\frac{1.04(1+\lambda)}{\lambda-\mu^*(1+0.62\lambda)}\right],
\end{equation}
}

\noindent where $\omega_{ln}$ is the logarithmically averaged phonon frequency, $\lambda$ is the electron-phonon coupling constant, and $\mu^*$ is the Coulomb pseudopotential which is set to be 0.1 in the calculations. We evaluate the pressure-dependent electron-phonon coupling constant $\lambda$ and the superconducting transition temperature $T_c$ shown in Fig. \ref{Tc}. It is found that both $\lambda$ and $T_c$ decrease, while $\omega_{ln}$ increases with increasing pressure. {Several other compounds have similar behaviors, especially at high pressures~\cite{Xu_2017,Bin_2014,HLiu_2017}). } The electron-phonon coupling constant given by phonon calculations ranges from 0.44 to 0.57, with corresponding $T_c$ from 2.5 K to 5.0 K. This result can be verified by further experimental studies at high pressure.

\section{Discussion and Conclusion}


When interacting electrons are spatially confined in the reduced dimensions, they show an enhanced tendency toward exotic electronic ground states, which makes them a fascinating topic of research, both theoretically and experimentally\cite{Haldane}. Q-1D systems provide an excellent playground in which to explore electronic correlations in low dimensions as they exhibit a rich variety of physical phenomena, including non-Fermi-liquid behaviors, superconductivity, charge density wave distortion and so on \cite{Haldane,Voit}. It is well established theoretically that the conventional picture of Fermi liquid theory does not apply when interacting electrons are confined to a single dimension, i.e. the Fermi liquid approach breaks down spectacularly in one-dimensional metals\cite{Giamarchi}. Instead, it can be described by the Tomonaga-Luttinger liquid paradigm, where the elementary excitations are collective spin and charge modes, propagating with different velocities and leading to spin-charge separation\cite{Giamarchi}. Surprisingly, for the q-1D molybdenum chalcogenides A$_2$Mo$_6$X$_6$, no experimental investigation has as yet been done in this respect, primarily due to the unfavorable sample dimensions of the single crystals available for the spectroscopic measurements, such as ARPES and STM. From the transport point of view, a Tomonaga-Luttinger liquid manifests a gross violation of an empirical law, i.e., the Wiedemann-Franz (WF) law which states that the ratio of the electronic thermal conductivity $\kappa_e$ to the electrical conductivity $\sigma$ at a given temperature $T$ is equal to a constant called the Lorenz number, $L_0 = \kappa_e / \sigma T =2.44 \times10^{-8} W\Omega/K^2$ and reflects that the same quasiparticles are responsible for both the thermal and electrical transport. This law was found to be strongly violated in the q-1D purple bronze Li$_{0.9}$Mo$_6$O$_{17}$ due to its spin-charge separation\cite{Xu09,Hussey11,Hussey19}. In the future, a similar test can be done for the q-1D A$_2$Mo$_6$X$_6$ to search for the possible non-Fermi liquid behaviors therein.

Quasi-one-dimensional conductors are prone to the Peierls distortion, forming the charge density modulation in real space. Indeed, some members in A$_2$Mo$_6$X$_6$ family, such as Na$_2$Mo$_6$Se$_6$, Rb$_2$Mo$_6$Se$_6$, Rb$_2$Mo$_6$Te$_6$ \textit{etc.}, undergo a metal-insulator transition at low temperatures, suggesting the CDW formation\cite{Hor85,Hor85_SSC}. For Tl$_2$Mo$_6$Se$_6$, there are two types of samples, one with metallic ground state while the other having resistivity upturn at low temperatures\cite{Armici1980}. This fact suggests that Tl$_2$Mo$_6$Se$_6$ is actually on the border of the Peierls transition, consistent with its q-1D Fermi surface and the soft phonon revealed in our calculations. Under a uniform pressure, our calculations suggest more three dimensionality of the electronic structure, in line with the experimental finding of the better metallicity at a small hydrostatic pressure\cite{Huang1983}. Interestingly, a uniaxial strain induces a metal-to-insulator transition in Tl$_2$Mo$_6$Se$_6$\cite{Tessema91}. The mechanism for this opposite trend under uniaxial and hydrostatic pressure merits a further study.

Tl$_2$Mo$_6$Se$_6$ also becomes superconducting below 3-5 K, depending on the sample quality. Very recently, Huang \textit{et al.} proposed that Tl$_2$Mo$_6$Se$_6$ is a time-reversal invariant topological superconductor induced by intersublattice pairing, favoring a spin-triplet order parameter with $E_{2u}$ symmetry\cite{Huang18}. Like topological superconductor Cu$_x$Bi$_2$Se$_3$, this odd-parity pairing would spontaneously break the rotational symmetry in its gap function and produce a nematic order\cite{Fu_liang10,Fu_liang14,Fu_liang16,Ando_Review,CuxBi2Se3_HC,CuxBi2Se3_NMR}. Certainly, it would be very interesting to probe the possible nematic superconductivity in Tl$_2$Mo$_6$Se$_6$ by, e.g., the high-resolution angle dependent calorimetric study\cite{CuxBi2Se3_HC}.

To summarize, via first-principles calculations, it is predicted that the topological properties of the q-1D conductor Tl$_2$Mo$_6$Se$_6$ change dramatically with pressure up to 50 GPa, above which a structural phase transition takes place to drive the system into a topologically trivial phase.
Our calculations also reveal the pressure dependence of the topological surface states and the phonon modes. Recently, it was predicted that, in the presence of strong interactions, CDSM or WSM can easily undergo a continuous quantum phase transition into either an axion insulator or a rotational symmetry-breaking nematic state\cite{Roy_PRB}. Our results establish Tl$_2$Mo$_6$Se$_6$ as an ideal arena for further exploring various topological phenomena associated with different types of topological fermions and may potentially be useful for engineering these nontrivial carriers in future applications.

\ack{
This work is supported by the National Natural Science Foundation of China (Grant No. 11674054, No. 11974061, No. U1732162, U1832147) and NUPTSF (Grant No. NY219087, NY220038). B. L. would also like to acknowledge the financial support from an open program from the National Lab of Solid State Microstructures of Nanjing University (Grant No. M32025).\\}




\begin{thebibliography}{0}
\bibitem{Hasan_RMP} Hasan M Z and Kane C L 2010 Colloquium: Topological insulators \emph{Rev. Mod. Phys.} \textbf{82} 3045
\bibitem{Zhang_RMP} Qi X L and Zhang S C 2011 Topological insulators and superconductors \emph{Rev. Mod. Phys.} \textbf{83} 1057
\bibitem{Zhou19} Zhou W, Li B, Xu C Q, van Delft M R, Chen Y G, Fan X C, Qian B, Hussey N E and Xu X F 2019 Nonsaturating Magnetoresistance and Nontrivial Band Topology of Type-II Weyl Semimetal NbIrTe$_4$ \emph{Adv. Electron. Mater.} 1900250

\bibitem{Bernevig16} Bradlyn B, Cano J, Wang Z, Vergniory M G, Felser C, Cava R J and Bernevig B A 2016 Beyond Dirac and Weyl fermions: Unconventional quasiparticles in conventional crystals \emph{Science} \textbf{353} 558
\bibitem{Zhang17} Tang P, Zhou Q and Zhang S C 2017 Multiple Types of Topological Fermions in Transition Metal Silicides \emph{Phys. Rev. Lett.} \textbf{119} 206402
\bibitem{Gibson2015} Gibson Q D, Schoop L M, Muechler L, Xie L S, Hirschberger M, Ong N P, Car R and Cava R J 2015 Three-dimensional Dirac semimetals: Design principles and predictions of new materials \emph{Phys. Rev. B} \textbf{91} 205128
\bibitem{Liu2017} Liu Q and Zunger A 2017 Predicted Realization of Cubic Dirac Fermion in Quasi-One-Dimensional Transition-Metal Monochalcogenides \emph{Phys. Rev. X} \textbf{7} 021019

\bibitem{Potel1980} Potel M, Chevrel R, Sergent M, Armici J C, Decroux M and Fischer ${\O}$ 1980 New pseudo-one-dimensional metals M$_2$Mo$_6$Se$_6$ (M = Na, In, K, TI), M$_2$Mo$_6$S$_6$ (M = K, Rb, Cs), M$_2$Mo$_6$Te$_6$ (M = In, TI) \emph{Journal of Solid State Chemistry} \textbf{35} 286
\bibitem{Armici1980} Armici J, Decroux M, Fischer ${\O}$, Potel M, Chevrel R and Sergent M {1980} A New Pseudo- One-Dimensional Superconductor Tl$_2$Mo$_6$S$_6$ \emph{Solid State Commun.} \textbf{33} 607
\bibitem{Huang1983} Huang S Z, Mayerle J J, Greene R L, Wu M K and Chu C W {1983} Pressure dependence of superconductivity in the pseudo-one-dimensional compound Tl$_2$Mo$_6$Se$_6$ \emph{Solid State Commun.} \textbf{48} 749
\bibitem{Tarascon1984} Tarascon J M, DiSalvo F J and Waszczak J V {1984} Physical properties of several M$_2$Mo$_6$X$_6$ compounds (M = group IA metal; X = Se, Te) \emph{Solid State Commun.} \textbf{52} 227
\bibitem{Hor85} Hor P H, Meng R L, Chu C W, Tarascon J M and Wu M K {1985} High pressure study on quasi-one-dimensional compounds M$_2$Mo$_6$X$_6$ \emph{Physica B} \textbf{135} 245
\bibitem{Hor85_SSC} Hor P H, Fan W C, Chou L S, Mengt R L, Chu C W, Tarascon J M and Wu M K {1985} Study of the metal-semiconductor transition in Rb$_2$Mo$_6$Se$_6$, Rb$_2$Mo$_6$Te$_6$ and Cs$_2$Mo$_6$Te$_6$ under pressures \emph{Solid State Commun.} \textbf{55} 231
\bibitem{Brusetti1988} Brusetti R, Monceau P, Potel M, Gougeon P and Sergent M {1988} The Exotic Superconductor Tl$_2$Mo$_6$Se$_6$ Investigated by Low Field Magnetization Measurements \emph{Solid State Commun.} \textbf{66} 181
\bibitem{Tessema91} Tessema G X, Tseng Y T, Skove M J, Stillwell E P, Brusetti R, Monceau P, Potel M and Gougeon P {1991} Probing the electronic structure in M$_2$Mo$_6$Se$_6$ \emph{Phys. Rev. B} \textbf{43} 3434
\bibitem{Brusetti94} Brusetti R, Briggs A, Laborde O, Potel M and Gougeon P {1994} Superconducting and dielectric instabilities in Tl$_2$Mo$_6$Se$_6$: Unusual transport properties and unsaturating critical field \emph{Phys. Rev. B} \textbf{49} 8931
\bibitem{Petrovic2010} Petrovi$\acute{c}$ A P, Lortz R, Santi G, Decroux M, Monnard H, Fischer {\o}, Boeri L, Andersen O K, Kortus J, Salloum D, Gougeon P and Potel M {2010} Phonon mode spectroscopy, electron-phonon coupling, and the metal-insulator transition in quasi-one-dimensional M$_2$Mo$_6$Se$_6$ \emph{Phys. Rev. B} \textbf{82} 235128

\bibitem{Giamarchi_2004} Giamarchi T {2004} Theoretical Framework for Quasi-One Dimensional Systems \emph{Chem. Rev.} \textbf{104} 5037


\bibitem{Petrovic2016} Petrovi$\acute{c}$ A P, Ansermet D, Chernyshov D, Hoesch M, Salloum D, Gougeon P, Potel M, Boeri L and Panagopoulos C {2016} A disorder-enhanced quasi-one-dimensional superconductor \emph{Nat. Commun.} \textbf{7} 12262
\bibitem{Gannon18} Gannon L, Boeri L, Howard C A, Gougeon P, Gall P, Potel M, Salloum D, Petrovi$\acute{c}$ A P and Hoesch M {2018} Lattice dynamics of the cluster chain compounds M$_{2}$Mo$_6$Se$_6$ (M = K, Rb, Cs, In, and Tl) \emph{Phys. Rev. B} \textbf{98} 014104
\bibitem{Chia18} Mitra S, Petrovi$\acute{c}$ A P, Salloum D, Gougeon P, Potel M, Zhu J X, Panagopoulos C and Chia E E M {2018} Dimensional crossover in the quasi-one-dimensional superconductor Tl$_{2-x}$Mo$_6$Se$_6$ \emph{Phys. Rev. B} \textbf{98} 054507
\bibitem{Huang18} Huang S M, Hsu C H, Xu S Y, Lee C C, Shiau S Y, Lin H and Bansil A {2018} Topological superconductor in quasi-one-dimensional Tl$_{2-x}$Mo$_6$Se$_6$ \emph{Phys. Rev. B} \textbf{97} 014510

{
\bibitem{Downing_2019} Downing C A, Sturges T J, Weick G, Stobi$\acute{n}$ska M, and Mart$\acute{i}$n-Moreno L {2019} Topological Phases of Polaritons in a Cavity Waveguide \emph{Phys. Rev. Lett.} \textbf{123} 217401
\bibitem{Pirmoradian_2018} Pirmoradian F, Rameshti B Z, Miri M and Saeidian S {2018} Topological magnon modes in a chain of magnetic spheres \emph{Phys. Rev. B} \textbf{98} 224409
}

\bibitem{Wien2k} Schwarz K, Blaha P and Madsen G K H {2002} Electronic structure calculations of solids using the WIEN2k package for material sciences \emph{Comput Phys Commun.} \textbf{147(1-2)} 71
\bibitem{GGA} Perdew J P, Burke K and Ernzerhof M {1996} Generalized Gradient Approximation Made Simple \emph{Phys. Rev. Lett.} \textbf{77} 3865
\bibitem{wannier1} Mostofi A A, Yates J R, Pizzi G, Lee Y S, Souza I, Vanderbilt D and Marzari N {2014} An updated version of wannier90: A tool for obtaining maximally-localised Wannier functions \emph{Comput. Phys. Commun.} \textbf{185} 2309
\bibitem{wannier2} Wu Q S, Zhang S N, Song H F, Troyer M and Soluyanov A A {2018} WannierTools: An open-source software package for novel topological materials \emph{Computer Physics Communications} \textbf{224} 405-416
\bibitem{QE} Giannozzi P \emph{et al.} {2009} QUANTUM ESPRESSO: a modular and open-source software project for quantum simulations of materials \emph{J. Phys.: Condens. Matter} \textbf{21} 395502
\bibitem{ux1} Oganov A R and Glass C W {2006} Crystal structure prediction using ab initio evolutionary techniques: Principles and applications \emph{J. Chem. Phys.} \textbf{124} 244704
\bibitem{ux2} Lyakhov A O, Oganov A R, Stokes H T and Zhu Q {2013} New developments in evolutionary structure prediction algorithm USPEX \emph{Comput. Phys. Commun.} \textbf{184} 1172
\bibitem{ux3} Bushlanov P V, Blatov V A and Oganov A {2019} Topology-based crystal structure generator \emph{Comput. Phys. Commun.} \textbf{236} 1-7

\bibitem{SKEAF} Rourke P M C and Julian S R {2012} Numerical extraction of de Haas-van Alphen frequencies from calculated band energies \emph{Comput. Phys. Commun.} \textbf{183} 324-32
\bibitem{Peierls} Peierls R E {1955} Quantum Theory of Solids (Oxford Univ Press, London)

\bibitem{Potter_NatComm} Potter A C, Kimchi I and Vishwanath A {2014} Quantum oscillations from surface Fermi arcs in Weyl and Dirac semimetals \emph{Nat. Comm.} \textbf{5} 5161
\bibitem{Hasan_Science_Na3Bi} Xu S Y, Liu C, Kushwaha S K, Sankar R, Krizan J W, Belopolski I, Neupane M, Bian G, Alidoust N, Chang T R, Jeng H T, Huang C Y, Tsai W F, Lin H, Shibayev P P, Chou F C, Cava R J and  Hasan M Z {2015} Observation of Fermi arc surface states in a topological metal \emph{Science} \textbf{347} 294
\bibitem{Hasan_Science} Xu S Y, Belopolski I, Alidoust N, Neupane M, Bian G, Zhang C, Sankar R, Chang G, Yuan Z, Lee C C, Huang S M, Zheng H, Ma J, Sanchez D S, Wang B K, Bansil A, Chou F, Shibayev P P, Lin H, Jia S and Hasan M Z {2015} Discovery of a Weyl fermion semimetal and topological Fermi arcs \emph{Science} \textbf{349} 613
\bibitem{Huang_NatComm} Huang S M, Xu S Y, Belopolski I, Lee C C, Chang G, Wang B K, Alidoust N, Bian G, Neupane M, Zhang C, Jia S, Bansil A, Lin H and Hasan M Z {2015} A Weyl Fermion semimetal with surface Fermi arcs in the transition metal monopnictide TaAs class \emph{Nat. Comm.} \textbf{6} 7373
\bibitem{Ding_NatPhys} Lv B Q, Xu N, Weng H M, Ma J Z, Richard P, Huang X C, Zhao L X, Chen G F, Matt C E, Bisti F, Strocov V N, Mesot J, Fang Z, Dai X, Qian T, Shi M and Ding H {2015} Observation of Weyl nodes in TaAs \emph{Nat. Phys.} \textbf{6} 7373


\bibitem{Kargarian} Kargarian M, Randeria M and Lu Y M {2016} Are the Surface Fermi Arcs in Dirac Semimetals Topologically Protected? \emph{Proc. Natl. Acad. Sci. USA} \textbf{113} 8648
\bibitem{Mcmillan1968} Mcmillan W {1968} Transition Temperature of Strong-Coupled Superconductors \emph{Phys. Rev.} \textbf{167} 331
\bibitem{Allen1975} Allen P and Dynes R {1975} Transition temperature of strong-coupled superconductors reanalyzed \emph{Phys. Rev. B} \textbf{12} 905

{
\bibitem{Xu_2017} Xu C Q, Sankar R, Zhou W, Li B, Han Z D, Qian B, Dai J H, Cui H, Bangura A F, Chou F C, and Xu X F {2017} Topological phase transition under pressure in the topological nodal-line superconductor PbTaSe2 \emph{Phys. Rev. B} \textbf{96} 064528
\bibitem{Bin_2014} Li B, Huang G Q, Sun J and Xing Z W {2014} Novel structural phases and superconductivity of iridium telluride under high pressures \emph{Scientific Reports} \textbf{4} 6433
\bibitem{HLiu_2017} Liu H, Naumov I I, Hoffmann R, Ashcroft N W, and Hemley R J Potential high-T$_c$ superconducting lanthanum and yttrium hydrides at high pressure \emph{Proc. Natl. Acad. Sci. USA} \textbf{114} 6990
}

\bibitem{Haldane} Haldane F D M {1981} 'Luttinger liquid theory' of one-dimensional quantum fluids: I. Properties of the Luttinger model and their extension to the general 1D interacting spinless Fermi gas \emph{J. Phys. C: Solid State Phys.} \textbf{14} 2585
\bibitem{Voit} Voit J {1994} One-dimensional Fermi liquids \emph{Rep. Rog. Phys.} \textbf{57} 977
\bibitem{Giamarchi} Giamarchi T {2004} Theoretical Framework for Quasi-One Dimensional Systems \emph{Chem. Rev.} \textbf{104} 5037
\bibitem{Xu09} Xu X F, Bangura A F, Analytis J G, Fletcher J D, French M M J, Shannon N, He J, Zhang S, Mandrus D, Jin R and Hussey N E {2009} Directional Field-Induced Metallization of Quasi-One-Dimensional Li$_{0.9}$Mo$_6$O$_{17}$ \emph{Phys. Rev. Lett.} \textbf{102} 206602
\bibitem{Hussey11} Wakeham N, Bangura A F, Xu X F, Mercure J F, Greenblatt M and Hussey N E {2011} Gross violation of the Wiedemann-Franz law in a quasi-one-dimensional conductor \emph{Nat. Comm.} \textbf{2} 396
\bibitem{Hussey19} Lu J, Xu X F, Greenblatt M, Jin R, Tinnemans P, Licciardello S, van Delft M R, Buhot J, Chudzinski P and Hussey N E {2019} Emergence of a real-space symmetry axis in the magnetoresistance of the one-dimensional conductor Li$_{0.9}$Mo$_6$O$_{17}$ \emph{Sci. Adv.} \textbf{5} eaar8027

\bibitem{Fu_liang10} Fu L and Berg E {2010} Odd-Parity Topological Superconductors: Theory and Application to Cu$_x$Bi$_2$Se$_3$ \emph{Phys. Rev. Lett.} \textbf{105} 097001
\bibitem{Fu_liang14} Fu L {2014} Odd-parity topological superconductor with nematic order: Application to Cu$_x$Bi$_2$Se$_3$ \emph{Phys. Rev. B} \textbf{90} 100509(R)
\bibitem{Fu_liang16} Fu L {2016} Odd-parity superconductors with two-component order parameters: Nematic and chiral, full gap, and Majorana node \emph{Phys. Rev. B} \textbf{94} 180504(R)
\bibitem{Ando_Review} Sato M and Ando Y {2017} Topological superconductors: a review \emph{Rep. Prog. Phys.} \textbf{80} 076501
\bibitem{CuxBi2Se3_HC} Yonezawa S, Tajiri K, Nakata S, Nagai Y, Wang Z, Segawa K, Ando Y and Maeno Y {2017} Thermodynamic evidence for nematic superconductivity in Cu$_x$Bi$_2$Se$_3$ \emph{Nat. Phys.} \textbf{13} 123
\bibitem{CuxBi2Se3_NMR} Matano K, Kriener M, Segawa K, Ando Y and Zheng G {2017} Spin-rotation symmetry breaking in the superconducting state of Cu$_x$Bi$_2$Se$_3$ \emph{Nat. Phys.} \textbf{13} 123


\bibitem{Roy_PRB} Roy B, Goswami P an Juri$\check{c}$i$\acute{c}$ V {2017} Interacting Weyl fermions: Phases, phase transitions, and global phase diagram \emph{Phys. Rev. B} \textbf{95} 201102(R)


\end{thebibliography}
\end{document}